\documentclass[preprint,journal]{vgtc}       % preprint (journal style)

\ifpdf%                                % if we use pdflatex
  \pdfoutput=1\relax                   % create PDFs from pdfLaTeX
  \usepackage{graphicx}                % allow us to embed graphics files
  \DeclareGraphicsExtensions{.pdf,.png,.jpg,.jpeg} % for pdflatex we expect .pdf, .png, or .jpg files
\else%                                 % else we use pure latex
  \usepackage{graphicx}                % allow us to embed graphics files
  \DeclareGraphicsExtensions{.eps}     % for pure latex we expect eps files
\fi%

\graphicspath{{figures/}{pictures/}{images/}{./}} % where to search for the images

\usepackage{microtype}                 % use micro-typography (slightly more compact, better to read)
\PassOptionsToPackage{warn}{textcomp}  % to address font issues with \textrightarrow
\usepackage{textcomp}                  % use better special symbols
\usepackage{mathptmx}                  % use matching math font
\usepackage{times}                     % we use Times as the main font
\usepackage{cite}                      % needed to automatically sort the references
\usepackage{tabu}                      % only used for the table example
\usepackage{booktabs}                  % only used for the table example
\onlineid{1020}

\vgtccategory{Research}
\vgtcpapertype{algorithm/technique}

\title{Scale-Space Splatting: Reforming Spacetime for Cross-Scale Exploration of Integral Measures in Molecular Dynamics}

\author{Juraj P\'{a}lenik, Jan By\v{s}ka, Stefan Bruckner, and Helwig Hauser}
\authorfooter{
\item
 Juraj P\'{a}lenik is with University of Bergen. E-mail: juraj.palenik@uib.no.
\item
 Jan By\v{s}ka is with University of Bergen and Masaryk University. E-mail: jan.byska@gmail.com.
\item
 Stefan Bruckner is with University of Bergen. E-mail: stefan.bruckner@uib.no.
\item
 Helwig Hauser is with University of Bergen. E-mail: helwig.hauser@uib.no.
}

\shortauthortitle{P\'{a}lenik \MakeLowercase{\textit{et al.}}: Scale-Space Splatting}
\abstract{Understanding large amounts of spatiotemporal data from particle-based  simulations\blue{, such as molecular dynamics,} 
often relies on the computation and analysis of aggregate measures.  
These, however, by virtue of aggregation, hide structural information about the space/time localization of the studied phenomena. \blue{This leads to degenerate cases where the measures fail to capture distinct behaviour.}
In order to drill into these aggregate values, we propose a multi-scale visual exploration technique.
Our novel representation, based on partial domain aggregation, enables the construction of a continuous scale-space for discrete datasets and the simultaneous exploration of scales in both space and time.  
We link these two scale-spaces in a scale-space space-time cube and model linked views as orthogonal slices through this cube, thus enabling the rapid identification of spatio-temporal patterns at multiple scales. 
To demonstrate the effectiveness of our approach, \blue{we showcase an advanced exploration of a protein--ligand simulation}.
} % end of abstract


\CCScatlist{ % not used in journal version
 \CCScat{K.6.1}{Management of Computing and Information Systems}%
{Project and People Management}{Life Cycle};
 \CCScat{K.7.m}{The Computing Profession}{Miscellaneous}{Ethics}
}

\teaser{
  \centering
  \includegraphics[width=\linewidth]{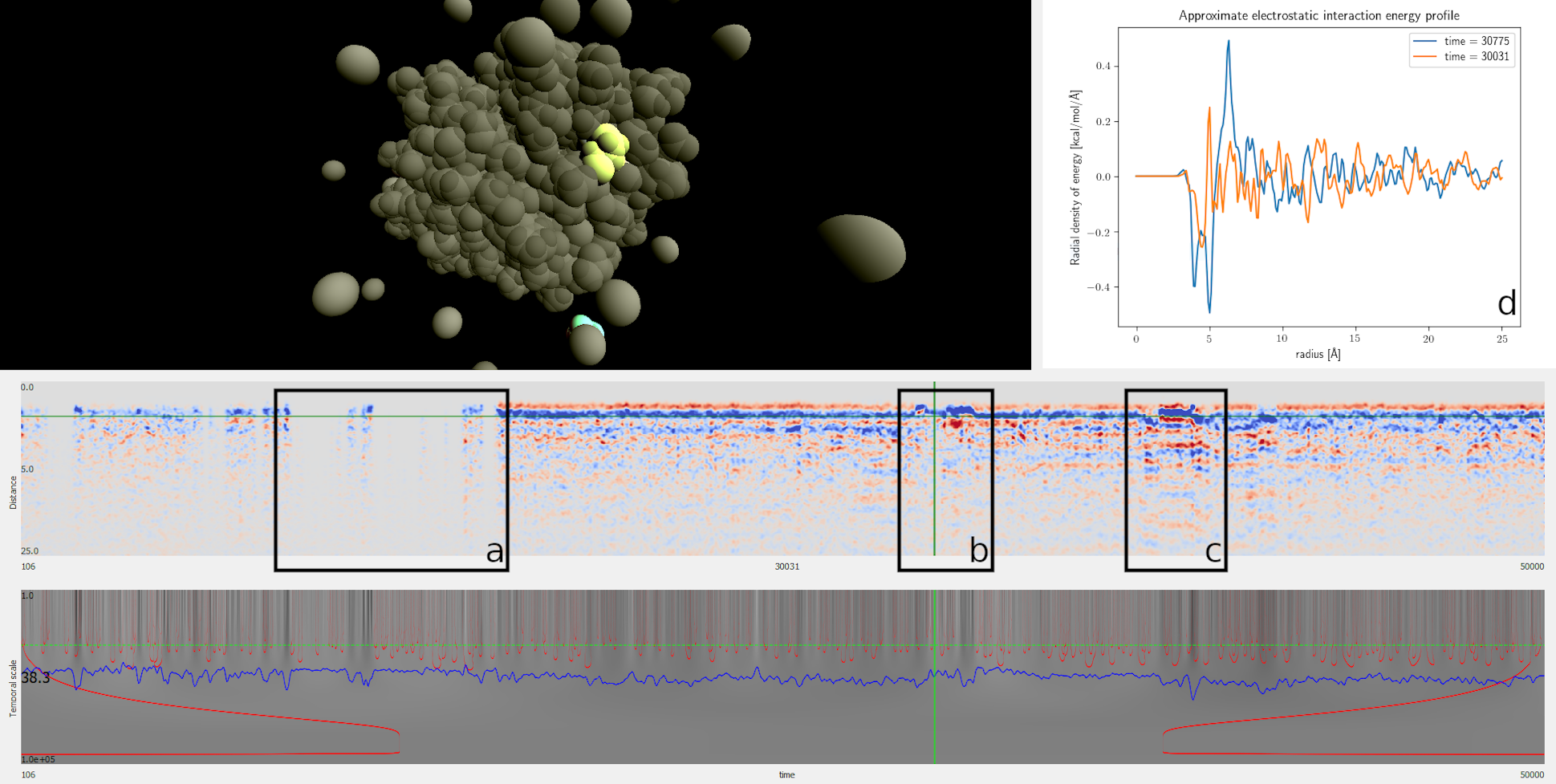}
  \caption{\label{fig:teaser}
    Novel representation of electrostatic interaction energy (in the middle) helps navigating the simulation by spatially resolving contributions to the integral measure traditionally represented by a timeseries (blue line in the bottom). It captures different modes of the simulation better than the timeseries and helps localizing changes in the configuration of the simulation by providing spatial context.
    The multiscale behaviour of the data is explored by seamlessly navigating the spatiotemporal scale-space. % efficiently computed by the Scale-Space Splatting method.
    The selected timestep shows the ligand escaping from the tunnel, before it is sucked back in by the protein.
  }%
}%

\vgtcinsertpkg
\usepackage{physics}
\usepackage{color}
\usepackage{float}
\usepackage{csquotes}
\usepackage{amsfonts}
\usepackage{epigraph}
\usepackage{subcaption}
\definecolor{darkgreen}{rgb}{0.0, 0.5, 0.16}

\newcommand{\red}[1]{}
\newcommand{\blue}[1]{#1}

\newcommand{\ssst}{scale-space space-time}
\newcommand{\gauss}{{g}}

\newcommand\restr[2]{{% we make the whole thing an ordinary symbol
  \left.\kern-\nulldelimiterspace % automatically resize the bar with \right
  #1 % the function
  \vphantom{\big|} % pretend it's a little taller at normal size
  \right|_{#2} % this is the delimiter
  }}

\begin{document}


\firstsection{Introduction}
%%%%%%%%%%%%%%%%%%%%%%%%%%%%%%%%%%%%%%%%%%
%%%%%%%%%%%%%%%%%%%%%%%%%%%%%%%%%%%%%%%%%%
\maketitle

%There are many domains, including physics, biology, chemistry, robotics, meteorology, and others, where the analysis of trajectory-based simulation data is a common practice.
% for gaining insight into relevant domain questions. 
The study of molecular dynamics (MD) simulations involves computation of descriptive integral measures such as energy, temperature, pressure, density, etc. % that are relevant for the analysis.  
These measures are, traditionally, computed over the length of the simulation and analysed as timeseries.
The purpose of such aggregations is to simplify the large amounts of data\blue{,  providing an overview of the simulation}, yet it comes at the cost of hiding possibly relevant structure.
\blue{This often results in degenerate timeseries that are not able to discriminate different modes in the simulation.}
%With this form of aggregation, spatial dependencies of the integral measures are not captured even though there usually is such a dependency. 
%After all, these aggregate measures are obtained as integrals (sums) over the spatial domain.  

In order to enhance the analysis of \blue{particle} simulations, \blue{we propose a novel representation that captures} spatial information of the aggregate measures.
Moreover, as the object of study is usually a large-scale dataset, the spatiotemporal dependency of the considered phenomena should be explored and analysed at multiple temporal and spatial scales.  
\par
For example, the electrostatic energy captures information about favourable and unfavourable states by summing contributions from spatially distributed electric charge.
However, all the information about the molecule's configuration in the spatial domain is lost by performing the spatial aggregation. 
Furthermore, the interaction energy time-series
captures various scales such as Brownian oscillation, amino-acid motion, and high-level changes, which all \blue{are} relevant for different types of analysis tasks. 
%This poses a challenge 

\red{Traditionally, specialized feature detection algorithms\cite{Lee2009}
%\cite{Midgaard, TimeZoom, FacetZoom, Wakame, MultiscaleBehaviour, Lee2009, Lichtenberg2018} 
are used to identify salient structures, but -- in particular when dealing with long time series -- it is often difficult to determine the appropriate scale of relevant features.} 
\blue{Traditionally, this complexity in the data is poorly supported by the current MD analysis programs. Advanced computational tools such as NAMD \cite{NAMD} support evaluation of the integral measures at predefined spatial and temporal resolutions, in practice however, the visual analysis \cite{CAVER3, VMD} defaults to time series representations, where the values are reconstructed at a single temporal scale, without regard for the spatial dimensions.}
\blue{Conversely,} scale-space approaches are well-established for investigating the behaviour of continuous functions across space and time \cite{Lindeberg1993, Lindeberg2011, Holmstrom2010, SiZer}. They provide valuable means for exploring global trends as well as local changes, and for 
treating noise in the data. 

In this paper, we present a novel approach to the simultaneous exploration of temporal and spatial scales of integral measures in MD data. In analogy to the projection of reconstruction kernels for splatting-based volume rendering \cite{Splatting}, we propose to project the Gaussian kernel used for scale-space construction into a reformed coordinate space that captures the space-time evolution of an integral measure. This enables us to provide a set of simple yet effective visual analysis mechanisms to explore the correspondences between the measure at different scales and its footprint in space and time.  

The main contributions of this work can be summarized as follows. We introduce a new approach to the visual analysis of trajectory data that specifically takes into account its multi-scale nature (in both time and space) and present scale-space splatting as a novel computational method for reconstructing a reformed space of integral measures. Furthermore, we show how this reformed space can provide insight into complex multi-scale phenomena, such as ligand--protein interaction, and demonstrate that it enables the visual identification of relevant features in real-world data. 

%%%%%%%%%%%%%%%%%%%%%%%%%%%%%%%%%%%%%%%%%%
%%%%%%%%%%%%%%%%%%%%%%%%%%%%%%%%%%%%%%%%%%
\section{Related Work}
%%%%%%%%%%%%%%%%%%%%%%%%%%%%%%%%%%%%%%%%%%

This paper contributes to the analysis of data that usually would be treated as a time series of spatially aggregated values.
A good overview of visualization techniques for time-dependent data can be found in a book by Aigner et al.~\cite{Aigner:2011:VTD}.  They focus primarily on time-dependent data in general, without specifically addressing spatial localization or scale-space approaches. 
Lee et al.~\cite{Lee2009} proposed an analysis based on multivariate trend identification in time dependent data without spatial dependence.
Wegenkittl et al. \cite{Wegenkittl1997} explore the use of parallel coordinates for the visualization of trajectories in multi-dimensional dynamical systems.

Our approach is based on the concept of the space-time cube, coined by H{\"a}gerstrand~\cite{hagerstrand1970space-time-cube} in 1970. Since then it was repeatedly used, either explicitly or as an underlying concept. 
Recently, Bach et al.~\cite{bach2017space-time-cube} presented a useful overview of related techniques. 
They describe the theoretical concept of a generalized space-time cube together with a taxonomy of all elementary space-time operations and their combinations that can be performed on such a space-time cube. 
%\\\indent
A more abstract approach to time-dependent volume data analysis has been proposed by Woodring et al. \cite{Woodring2003}, based on hyperplane slicing of a four-dimensional space-time hypercube.
Other examples of slicing higher-dimensional data (not necessarily time dependent) include Sliceplorer~\cite{Sliceplorer}, Hyperslice~\cite{Hyperslice}, Hypersliceplorer~\cite{Hypersliceplorer} and HyperMoVal\cite{HyperMoVal}.
%
%%%%%%%%%%%%%%%%%%%%%%%%%%%%%%%%%%%%%%%%%%
\subsection{Scale-Space Construction}
%%%%%%%%%%%%%%%%%%%%%%%%%%%%%%%%%%%%%%%%%%
%
The idea of a scale-space representation was first suggested by Witkin~\cite{Witkin1983}.
He proposed an approach to constructing a scale space by continuous blurring of the original time series with a Gaussian kernel of increasing size.
In his work, the scale-space is \red{then }segmented \red{in regions of monotonicity }by finding the zero-crossings of the second derivative, and an interval tree is constructed and visualized as a tesselation of the scale-space.
\par
In the image processing field, a well-known generalization of the scale-space construction has been proposed by Perona and Malik~\cite{PeronaMalik}, who slackened the uniformity constraint and built the scale-space using anisotropic diffusion.
Although their designated use-case was edge detection in image data, the method itself has been successfully applied to time-series scale-space construction as well~\cite{Lin08:thesis}.
%\par
A spatiotemporal extension of the scale-space approach has been presented by Laptev~\cite{Laptev2005}. He extended the space for 2D time-dependent data and his work has been successfully applied to feature detection in videos.
For more advanced scale-space solutions, including an $n$-dimensional time-dependent scalar field scale-space, we refer the reader to Tony Lindeberg's book \enquote{Scale-Space Theory in Computer Vision} and his more recent works~\cite{Lindeberg1993, Lindeberg2011}.
%\\\indent
The respective works on a spatiotemporal scale-space construction by Laptev and Lindeberg are unfortunately not directly applicable to particle data. 
%Moreover it is mainly focused on the scale-space construction and feature detection for image processing, instead of visualization.
%\red{Do you guys know of any direct visualization techniques for the spatiotemporal scale-space of Tony Lindeberg?}
\par
Meijers and Oosterom~\cite{meijers2011space-scale-cube} described a scale-space construction for polygonal maps, which they named \emph{space-scale cube}. 
They describe a hierarchical level-of-detail method for geographical chart construction, but are not concerned with temporal data.  
%\blue{I have the impression that the overlap is only in the name. They compute discrete LOD (level of detail) topological structures. They name the work scale-space cube, however, they are more concerned with the construction and application than introducing a concept of scale-space exploration.}
\par
More scale-space techniques can be found in an overview by Holmstr\"{o}m \cite{Holmstrom2010}.
Vuollo et al. \cite{VUOLLO2018} present the construction of a scale-space for spherical data. In their work the data populates the surface of a sphere which is topologically different from our reformed coordinates.
%

%
%
%%%%%%%%%%%%%%%%%%%%%%%%%%%%%%%%%%%%%%%%%%
\subsection{Scale-Space Applications in Visualization}
%%%%%%%%%%%%%%%%%%%%%%%%%%%%%%%%%%%%%%%%%%
% 
%Hao et al. \cite{Hao2007} proposed pixel efficient representation of long time series with importance driven resampling 
%
Scale-space approaches have been successfully applied in the field of visualization. 
%\todo{perhaps there are many more papers.}
%\\\indent
For example, a scale-space surface extraction for 3D density fields was proposed by Kindlmann et al.~\cite{Kindlmann2009}. They utilize a precise scale interpolation technique for the detection of creases by an energy-function driven particle sampling. They construct the scale-space for a continuous, static 3D field and sample the scale-space with particles as discrete agents, but they do not construct a scale-space for particle data.
\par
Klein and Ertl~\cite{Klein07} describe the scale-space tracking of critical points in 3D vector fields. Their technique is based on 3D continuous static data.
%\\\indent
Miao et al.~\cite{Miao:2019:MMV} address the multi-scale characteristics of large molecular simulations (an HIV virus modelled at the atomic level). The authors are, however, concerned with the multi-scale nature of the spatial domain only and do not address the temporal aspect of the simulation. 
%\par
Bremer et al.~\cite{bremer:hal} proposed a method for spatial scale selection in grid-based simulations, founded on Morse theory. 
In their work, a tree decomposition of the spatial domain is constructed, and a visual exploration of threshold values is enabled as opposed to a priori value guessing. 
The method is based on a grid representation and thus not directly applicable to particle simulations. 
\par
Pinus, by Sips et al.~\cite{pinus}, is a multiscale visual analytics technique for finding patterns in time series data. 
In this work, a tree representation is constructed by computing hierarchical aggregates of the time series. 
While this approach is applicable to time series data, their paper is not concerned with the spatial aspect.

\subsection{Space Reformations}
Space reformation techniques are valuable tools for gaining insight into the spatial structure of the data.
%\\\indent
Recent publications present methods for a decomposition using a space-filling curve in MotionRugs~\cite{MotionRugs} and Dynamic Volume Lines~\cite{Weissenbock2018}. 
These works transform continuous volume data into one-dimensional representations by following a space-filling curve through the volume.
As this kind of space reformation suffers from \enquote{delocalization}, where points close to each other can end up far apart in the representation, Dafner et al.~\cite{Dafner2000} tried to overcome this limitation by context-based space-filling curves. 
A space-filling curve is not applicable to our case as it does not provide the desired aggregation required for the statistical treatment of particle data.
\par
%For flow data Angelelli et al.\cite{Angelelli2011} devised a space transformation for straightening tubular flows. 
Several works address space reformation techniques for volume data, of which we highlight two:
A general approach to volume transformation by Chen et al.~\cite{Chen2003}, via the use of spatial transfer functions for 3D volume warping, and a curved planar reformation by Kanitsar et al.~\cite{Kanitsar2003}, that was used in medical visualization.
For more examples, Kreiser et al. \cite{FlatteningSurvey} provide a survey of flattening based techniques in medical visualization.
%Depending on which case is considered, 
%\blue{Our approach is based on spatial symmetries, however, different space reformation methods based on domain integration  to incorporate in our approach.}
% These transformations are very domain-specific, and one has to tailor them for the particular use so we don't describe these in detail.
% However, such transformations might be, in theory, integrated into our method as volume preprocessing operations.
%
\subsection{Particle Simulation Data}
Substantial work has been done on the visualization of trajectory-based simulation data, including projects such as OVITO~\cite{OVITO}, Trillion Particles~\cite{TrilionParticles}, Multiscale HIV~\cite{Miao:2019:MMV}. These are primarily concerned with the sheer volume of the data and its direct visualization, rather than the computation of derived properties and their temporal analysis.

Focusing on the trajectories, hierarchical particle grouping for large datasets has been done by Fraedrich et al. \cite{Fraedrich2012}. Schirski et al. \cite{Schirski2004} extract prominent trajectories from large particle data.
Kottravel et al. present a specialized tool for exploring Monte Carlo simulations of photo-voltaic cells.

\par
Lichtenberg et al.~\cite{Lichtenberg2018} present a visual analytics tool for exploring molecular structures based on the Solvent Accessible Surface (SAS).
This is a geometric approach to the extraction of the spatial configuration of a protein--ligand interaction.
%\\\indent
MoleCollar and Tunnel Heat Map \cite{MoleCollar} are works by By\v{s}ka et al. where they reform the properties of a protein tunnel into the tunnel's centre line and the tunnel's cross section.
In AnimoAminoMiner \cite{AminoMiner}, the authors focus on amino acids lining the tunnel and their temporal development.

A visual analytics tool for long molecular dynamics simulation data, described by Duran et al.~\cite{Duran2018}, provides an importance-driven time series analysis of integral measures of particle simulation data.
They rely on time series aggregation and interval clustering based on ligand's position.
This approach is not concerned with resolving the spatial dependency of the integral measures.
VIA-MD \cite{molva.20181102} uses histograms to capture the temporal structure of complex spatio-temporal data from molecular dynamics and volume rendering to show the spatial probability distribution. 

%\blue{None of the approaches above tries to directly visualize the spatial structure of the integral measures at multiple scales.}
Our approach is complementary to those above and focuses on resolving the spatial dependency of the integral measures for the cases when the timeseries representation fails to capture the relevant changes.

%\subsection{High dimensional exploration}
%
%Sliceplorer \cite{Sliceplorer}
%Hyperslice \cite{Hyperslice}
%Hypersliceplorer \cite{Hypersliceplorer}
%HyperMoVal \cite{Hypermoval}

%
%
%%%%%%%%%%%%%%%%%%%%%%%%%%%%%%%%%%%%%%%%%%
%%%%%%%%%%%%%%%%%%%%%%%%%%%%%%%%%%%%%%%%%%
\section{On Scale-Space Reformation}
%%%%%%%%%%%%%%%%%%%%%%%%%%%%%%%%%%%%%%%%%%
%%%%%%%%%%%%%%%%%%%%%%%%%%%%%%%%%%%%%%%%%%
%
We see that large-scale and long-time particle simulations are common in a variety of fields, especially in molecular dynamics (MD).
The exploration of both the spatial configuration and aggregate measures is of importance to domain experts.
The traditional representation of an aggregate measure as a time series, used for its simplifying character, cannot convey any spatial configuration.
Furthermore, it exhibits multiscale behaviour that is usually not explicitly addressed.
Scale-space approaches are used in continuous field analysis for the treatment of multiple scales in the data, however, they are expensive to compute for large datasets and they are not easily combined with spatial reformations that could bring out the most important spatial dependencies.
Enabling an efficient computation of scale-space in reformed coordinates can significantly speed up the algorithms and allow for simplified visualizations exploiting dimensionality reduction.
In this section, we address these issues and describe the construction of a scale-spaces in reformed coordinates.

%
%\red{As there is no task abstraction in the introduction of Continuous Scatterplots, there is no method justification, the section starts with defining the mathematical model.}
%
%In this section we introduce the method, starting with an overview and then diving into the details.
%\todo{what? and why?}

The canonical scale-space is defined as the evolution of a diffusion equation over a continuous field \cite{Lindeberg2011}.
The standard way to recover a continuous field from particle data (for example in Smoothed Particle Hydrodynamics) is by kernel density estimation \cite{sph}.
In this approach, each particle is replaced by a kernel function, centred at the particle position, and the contributions from all the particles are summed up into a single scalar field. 
The size of the reconstruction kernel affects the resulting function. 
The bigger the kernel used, the smoother the resulting function becomes, with less detailed features.

One can obtain a scale-space of the data by evolving the reconstructed field with the diffusion equation.
The solution to the diffusion equation can be obtained by convolving the field with a Gaussian kernel.
If the reconstruction kernel is also a Gaussian, the same scale space can be obtained by a series of reconstructions with increasing kernel size.
%This holds, because a convolution of two Gaussians is again a Gaussian, which can be easily proven by a Fourier transformation of the convolution.
This gives a straightforward and easy to grasp concept of a scale-space for particle data. 
By gradually increasing the size of the reconstruction kernel we achieve the same results as by using the convolution on the reconstructed field. 
%Both of which can be used for efficient implementations.

Adapting this to time-dependent data means stepping up to four-dimensional space-time and evolving the diffusion equation in time, as well. 
As Lindeberg et al. \cite{Lindeberg2011} explain, the size of the kernel in the spatial dimensions does not behave the same as the one along the temporal dimension.
In their work on video data, they managed to bind the two scale dimensions which enabled a joint extrema analysis and an automatic identification of space-time points of interest. 
Finding such a formula is application dependent and restricts the flexibility of the analysis. % analytical powers of the scale-space.
Therefore, we treat the two scales separately, ending up with two independent scale dimensions, which we address by the means of visual exploration.

Our scale-space construction leaves us with a six-dimensional continuous scalar field with three spatial dimensions, a single time dimension and two scale dimensions. 
High dimensional slicing techniques could be utilized to explore such data, for example Sliceplorer~\cite{Sliceplorer} or Hyperslice \cite{Hyperslice}.
%Trying to replace a timeseries (1D data) with a 6D scale-space space-time field representation as a simple overview and navigation tool is ridiculous. 
%\red{This is just a bomb spec, coming from nowhere. Needs to go to intro.} 
%
However, constructing this six-dimensional space means engaging with an enormous analytical challenge, while it is not always necessary. 
We find that often the explored phenomena are distance-based and/or come with a spatial symmetry that can be exploited to reduce the spatial complexity of the system. 
%In layman's terms: often it is enough to know how far away from a reference point the particles are. 

The analysis with respect to reference structures leads us to three standard transformations of coordinates, known from calculus.
In the first approximation, these reference structures are a point, a line and a plane.
For example, we might want to study a molecular dynamics simulation with respect to the position of a ligand interacting with the main molecular structure (e.g; a protein).
Alternatively, a probe in a plasma physics scenario is usually formed by a straight wire, immersed in ionized matter, which can be considered as a line of interest/reference.
Further, the distance from the piston in an engine combustion simulation can define a reference plane of interest.

In order to analyse these examples, one would simplify their spatial dependency by projecting the particle positions onto reformed coordinates. 
Since computing the scale-space in three dimensions in high resolution is a challenge, especially for large datasets, our approach aims at recognizing useful symmetries and realizing an according lower-dimensional scale-space in the reformed coordinates.
This amounts to projecting the Gaussian kernel onto the reformed axes and using the projected version to convolve the data \blue{in the lower dimensional space}.

\subsection{Scale-space basics}\label{sec:scale-space}
A scale-space is usually a higher dimensional space constructed from an original field by promoting a reconstruction-filter parameter to a new continuous dimension.
In the original idea of a scale-space described by Witkin \cite{Witkin1983}, a one-dimensional input function $f(t)$ is filtered with a Gaussian blurring kernel with increasing kernel size, creating a two-dimensional scale-space ``image'' $G(t,\tau)$:
\begin{equation}\label{eq:temporal_scale}
G(t,\tau) = \int_{-\infty}^\infty f(u) \frac{1}{\sqrt{2\pi \tau^2}} \exp \left[ {{-\frac{(t-u)^2}{2\tau^2}}} \right] \dd u
\end{equation}
with $\tau$ being the kernel size. Note that\blue{, throughout the paper,} we use $\sigma$ for the spatial kernel \blue{size} and $\tau$ for the temporal kernel size.

In order to detect and trace important features across scales, level sets of scale-space derivatives can be used.
Originally, the zero crossings of the second derivative were proposed to identify peaks and valleys:
\begin{equation}\label{eq:contours}
\pdv[2]{G(t,\tau)}{t} = 0
\end{equation}
In the case of a one-dimensional scale-space this leads to one dimension for time and one dimension for the scale. 
The zero crossings of the second derivative then yield one-dimensional contours, tracing the time location of the inflexion point over the scales.
%The more prominent a feature, the bigger the contour along the scale axis. 
%

%A more modern definition~\cite{Lindeberg2011} states that the scale-space is obtained as the time evolution of the diffusion equation
%\begin{equation}
%\pdv{G(\tau, \vec{x})}{\tau} = \nabla_x \cdot (\mathbf{C}(\tau,\vec{x}) \cdot \nabla_x G(\tau, \vec{x})),\quad \tau \in \mathbb{R}^+
%\end{equation}
%with the boundary condition $G(0, \vec{x}) = f(\vec{x})$. Where $\mathbf{C}$ is the diffusivity tensor, $f(\vec{x})$ is the input field and $\tau$ corresponds to the scale parameter.
%Solving this differential equation with constant diffusivity using the method of impulse response yields the Gaussian convolution.

\subsection{Definitions}
In this paper, we treat MD simulation as trajectory-based particle data together with a descriptive integral measure.
By an integral measure, we understand any function that can be computed from the trajectory data for each time-frame as a sum over all particles \blue{(atoms)}.
Examples are: energy, potential, temperature, pressure, density, and more.  

The particles ``live'' in the spatial domain, denoted $\Omega \subseteq \mathbb{R}^3$.
By considering each particle in its own copy of $\Omega$, we get a Cartesian pro\-duct of $N$ copies representing a spatial configuration of N particles:~$\Omega^N$.
We denote the integral measure as a function $F:\Omega^N \rightarrow \mathbb{R}$, defined as:
\begin{equation} 
\begin{aligned}\label{eq:state_fnc}
%F& : \Omega^N \rightarrow \mathbb{R}\\\
F(\vec{x}_1(t),\dots,\vec{x}_N(t))& = \sum_{i=1}^N \mathcal{F}_i\,(\vec{x}_i(t)) 
\end{aligned} 
\end{equation}
where $\Omega^N$ is the configuration space, $\mathcal{F}_i$ is a function evaluated per particle with position $\vec{x}_i(t)$ at the time $t$ and $i$ is indexing the particles. 
\blue{
The first requirement is, therefore, that the integral measure is computed as a sum of per particle contributions.
}

\blue{The traditional representation of the integral measure, by a timeseries $f(t)$, is obtained by summing all the contributions for each timestep.} 
\begin{equation}    \label{eq:timeseries}
    f(t) = F(\omega(t)),
\end{equation}
where $\omega(t) \in \Omega^N$ is a particular spatial configuration of the particle system at time $t$. 
\blue{For example, evaluating the energy for each timestep yields the an energy timeseries.}

\blue{A continuous representation of the integral measure capturing the spatial information can be obtained from particle data by kernel density estimation (KDE)~\cite{KDE-1, KDE-2, KDE-3}}. The KDE with kernel size $\sigma$ yields a smooth scalar valued function $\Phi_\sigma(t, \vec{x}): \mathbb{R} \times \Omega \rightarrow \mathbb{R}$
\begin{equation}\label{eq:Phi_def}
\Phi_\sigma(t, \vec{x}) = %\int_\Omega \dd \vec{x}\ 
%\left(
%\underbrace {
\sum_{i=1}^N  \mathcal{F}_i\,(\vec{p_i}(t))\ \gauss_\sigma(\vec{x} - \vec{p}_i (t))
%}_{}
%\right)
\end{equation}
where $\gauss_\sigma$ is the kernel function with kernel width $\sigma$ and $\mathcal{F}_i(\vec{p}_i)$ is the contribution from the particle at the position $\vec{p}_i$. 
\blue{In case of the energy, this construction yields the energy density function. The use of Gaussian kernel provides us with two important properties described below.}
\red{
There are two ways of computing the KDE. 
The first one, as denoted in Eq.~\ref{eq:Phi_def}, where the contribution of a particle is evaluated at its position and the result is smeared over the neighbourhood by a Gaussian kernel. 
The second one can be obtained by using $\mathcal{F}_i(\vec{x})$ instead of $\mathcal{F}_i(\vec{p}_i)$ in the Eq.~\ref{eq:Phi_def}. 
The particle is again replaced by a kernel density function, but now the contributions are re-evaluated at all points in space and the respective exact contributions weighted by the kernel.
This results in a more accurate reconstruction, but it is also much more computationally expensive.
To cope better with larger data, we employ the first one, also for its nicer analytical properties as described below.
\begin{flushright}
[This paragraph was removed, because it does not explain a part of the method and complicates the section. The "second way" of computing the KDE is not utilized, and it is \textbf{not} more common in the literature due to a high computational cost.]
\end{flushright}
}

First, the convolution of a Gaussian kernel with the KDE yields a reconstruction with a bigger kernel:
\begin{equation}
\Phi_{\sigma+\tau}(t, \vec{x}) = \gauss_\tau(\vec{x}) \ast \Phi_\sigma(t, \vec{x})
\end{equation}
This allows us to \blue{avoid the a priori specification of a single reconstruction scale and instead} call the family of KDE reconstructions a scale-space with a positive, continuous scale parameter $\sigma \in \mathbb{R}^+$.

Second, integrating the KDE $\Phi(t,\vec{x})$ over the domain $\Omega$ yields the time series $f$:
\begin{equation} \label{eq:Phi_integral}
%\tilde{f}_\sigma(t) 
f(t) = \int_\Omega   \Phi_\sigma(t, \vec{x})\ \dd \vec{x}
\end{equation}
Since all the $\mathcal{F}_i(\vec{p}_i)$ contributions are constant \blue{at a fixed time $t$}, we just need to integrate over the normalized kernel. \blue{The kernel always integrates to one, resulting in the sum over all particle contributions, which is the timeseries $f$.}
This \blue{expedient} behaviour with respect to integration motivates the following section.
%These timeseries will in general slightly differ, although, they capture the same measure. In the limit of $\sigma \rightarrow\ 0$, both timeseries are identical. 
%\begin{equation}
%\lim _{\sigma \rightarrow\ 0} \tilde{f}_\sigma(t) = f(t)
%\end{equation}
%
%	There is a secret third property!
%
%Third, with $\sigma \rightarrow\ \infty$, the $\Phi_\sigma(t,\vec{x})$ will approach $ \frac{1}{V} f(t)$ in every point $\vec{x}$. Where $V$ is the volume of the domain $\Omega$.
%%%%%%%%%%%%%%%%%%%%%%%%%%%%%%%%%%%%%%%%%%

%%%%%%%%%%%%%%%%%%%%%%%%%%%%%%%%%%%%%%%%%%
\subsection{Space Reformation}\label{sec:reform}
%%%%%%%%%%%%%%%%%%%%%%%%%%%%%%%%%%%%%%%%%%
Our main goal is to understand the relevant dependencies of the integral measure $F$ on the spatial configuration of the particles. 
Since the function $F$ may describe different kinds of interaction between the particles (long-range interactions, short-range interactions, collective behaviour, etc.), \blue{different spatial structures will emerge}.
\par

%One might observe, that the original (not integrated) representation ${F}$ captures the spatial dependency exactly. As it is a scalar valued function of the configuration space $\Omega^N$ it cannot be visualized for more than a handful of particles.\todo{Did perhaps Helwig try to visualize the configuration space? Other ref?}
An initial attempt at exploring the spatial dependency would be through the kernel density estimate $\Phi_\sigma(t,\vec{x})$. This function could be reconstructed at multiple scales and explored by direct visualization techniques, using, for example, efficient SPH rendering \cite{SPH-Rendering} or implicit Gaussian surfaces \cite{Bruckner-2019-DVM}.
The full spatial reconstruction of this field, however, imposes big challenges on the analysis, where this is not necessarily required to capture the relevant spatial dependency of the studied integral measure.
Many measures can, for example, be meaningfully studied as distance-based, where only the relative distance to a reference structure is important. 

If we can capture this dependency with a model of the domain \blue{using a transformation of coordinates denoted} $\psi: \Gamma \rightarrow \Omega$, we can reform the coordinates, such that the relevant spatial information is contained in the first dimension of \blue{the reformed space} $\Gamma$. 
We can then project the KDE by a partial integration of the domain:
\begin{equation} \label{eq:projection}
\Phi_\sigma^\star(t, u) = \int_{\restr{\psi^{-1}(\Omega)}{v,w}} \hspace{-0.5em} \Phi_\sigma(t, \psi(u,v,w)) |\det(D\psi)(u, v, w)|\ \dd v\ \dd w
\end{equation}
\blue{A coordinate transformation of an integral is performed by substituting the new variables and including the determinant of the Jacobian matrix $|\det(D\psi)|$. The projection is then obtained by} integrating in the new coordinates $u, v, w$ over everything except the new, independent variable~$u$. 
\blue{This results in a simplified representation of the kernel density estimate, where the relevant spatial structure is captured in a single, new spatial dimension. For example, in case of a spherical transformation the new representation would be called a radial density.}

Looking at the definition of $\Phi$ in Eq.~\ref{eq:Phi_def}, it consists of a weighted sum of kernels.
Hence, just as in Eq.~\ref{eq:Phi_integral}, the projection onto the relevant axis can be computed using the projections of the kernels. 
%This approach is similar to the volume splatting algorithm by Lee Westover. \cite{Splatting}
The reconstructed KDE in the projected coordinates then boils down to
\begin{equation}\label{eq:Phi_star}
\Phi^\star_\sigma(t, u) = \sum_{i=1}^N  \mathcal{F}_i\,(\vec{p_i}(t))\ \gauss^\star_\sigma(u(\vec{x}), u(\vec{p}_i (t)))
\end{equation}
where $\gauss_\sigma^\star$ is the projected kernel and $u(\vec{x})$ is the coordinate $u$ of the transformation $\psi^{-1}: \Omega \rightarrow\ \Gamma$.

%Depending on the case and the according location of interest, also other more complicated models may be relevant.
We take the traditional integral transformations from calculus \cite{SpivakMichael1998} based on the dimensionality of the region of interest: 
%\begin{myitemize}
1) point of interest $\to$ spherical transformation;
2) line of interest $\to$ cylindrical transformation;
3) area of interest $\to$ orthogonal transformation.
%\end{myitemize}
%The kernel projections for these three basic cases are in the following subsections. 
%

\blue{The resulting projections of the Gaussian kernels can be seen in Fig.~\ref{fig:both} and the derivations in the supplementary material. Looking at Fig.~\ref{fig:both} we can observe that the projection onto both the spherical and cylindrical radial distance mostly takes care of the boundary condition when $r \rightarrow\ 0$ and behaves like a traditional Gaussian for $r\rightarrow\infty$.}

\begin{figure}[tbhp]
\includegraphics[width=\columnwidth]{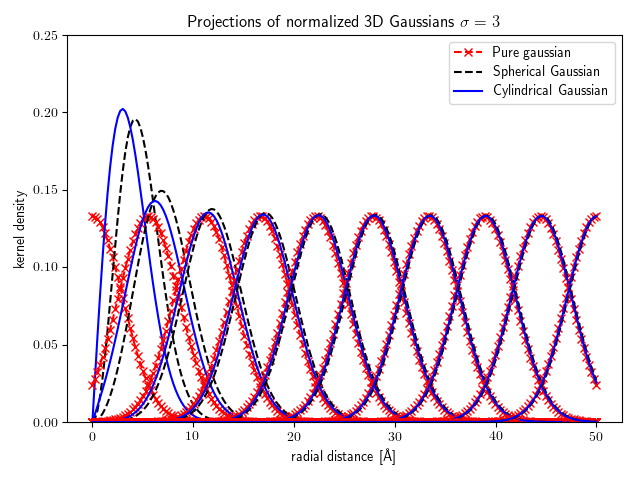}
\caption{\label{fig:both}
Spherically (black) and cylindrically (blue) projected Gaussians compared with pure Gaussian kernels (red crosses) centred at positions uniformly sampling the radial axis. 
%Each projected Gaussian is a product of a red Gaussian and green scaling function. 
We can observe that the projection takes care of the boundary condition when $r \rightarrow\ 0$. \blue{Each kernel that would extend into the negative domain ($r<0$) gets \enquote{pushed out} such that it zeros out at $r = 0$.}}
\end{figure}

\begin{figure*}[t!]
\centering
\includegraphics[width=0.85\linewidth]{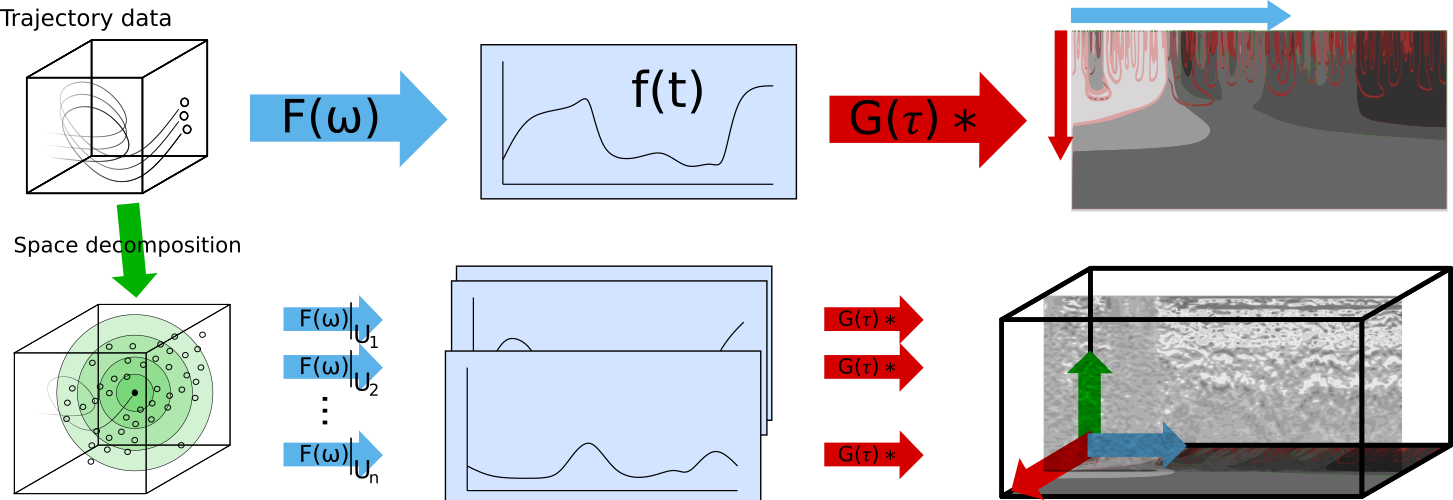}
\caption{Overview of the temporal scale-space construction. 
{\em{}First row}: the integral measure $F(\cdot)$ is evaluated for each frame of the particle data, yielding a time series $f(t)$, which is then convolved with a set of Gaussian kernels, resulting in a traditional scale-space; 
{\em{}Second row}: A point location of interest is identified; the spatial domain is decomposed into shells; the integral measure is evaluated for each region giving a set of time series; the scale-space construction is repeated for each and the results are stacked into the SSST cube. %The spatial scale is considered by varying the size of the shells.
}
\label{fig:schema}
\end{figure*}

\subsection{Time-Scale Representation}
%%%%%%%%%%%%%%%%%%%%%%%%%%%%%%%%%%%%%%%%%%
%\todo{Matter of fact style..}
Treatment of the temporal evolution is illustrated in Fig.~\ref{fig:schema}. 
\blue{First the traditional representation of the integral measure by a timeseries $f(t)$ is computed.
The timeseries describes the temporal evolution of the integral measure, disregarding the spatial structure.}
To capture the temporal scales in the timeseries the scale-space representation is generated as described in Sec.~\ref{sec:scale-space}, obtaining a scale-space image and a level set of zero crossing curves.
\blue{The scale-space image is used for the temporal scale exploration as detailed in Sec.~\ref{sec:scale-view}.}
This corresponds to the first row of Fig.~\ref{fig:schema}.

Constructing the temporal scale-space for the KDE representation\blue{, that captures the spatial structure, is done by repeating the very same process.}
Considering that the function $\Phi(t,\vec{x})$ \blue{(representing the KDE)} consists of a timeseries in each point $\vec{x}$, we apply Eq.~\ref{eq:temporal_scale} at each of these points.
This amounts to convolving the data along the temporal axis with Gaussian kernels of increasing sizes\blue{, as expected from treating the time and space independently.
This construction, however, creates a large amount of data that would necessarily be projected and for analytical purposes.} 
\blue{To avoid this large amounts of intermediate data, we construct the temporal scale-space of the projected KDE representation directly.}
Looking at Eq.~\ref{eq:projection}, we see that the projection is done by integrating over the spatial domain only.
Plugging the definition of $\Phi^\star(t, u)$ into Eq.~\ref{eq:temporal_scale} and using Fubini's theorem \cite{SpivakMichael1998} to exchange the order of the integrals, we prove that the scale-space construction and the space reformation are order independent. 
\blue{The projected temporal scale-space is therefore obtained by a direct convolution of $\Phi^\star(t, u)$ with Gaussian kernels along the temporal axis.}

%We should be careful about this result. 
Even though the order of the operations is not important, the transformation itself will introduce temporal dependency if it changes over time.
It is important to keep this mind when interpreting the results, however, it is still a satisfying result as it enables us to first project the data and construct the scale-space of much smaller representation.
\blue{The temporal dependency introduced by the projection will be the same as the one introduced by projecting the temporal scale-space of the full KDE. The correct choice of a space reformation will capture the spatial dependency in the data.  Whereas the scale-space representation helps identifying features at different scales.
The scale-space in the reformed coordinates is the basis for the slice view detailed in Sec.~\ref{sec:slice-view}.}

The construction of the temporal scale-space in the reformed coordinates is illustrated in the second row of Fig.~\ref{fig:schema}.

%The integral measure $F(\cdot)$ is evaluated for each frame of the particle data, yielding a time series $f(t)$, which is then convolved with a set of Gaussian kernels, resulting in a traditional scale-space.
%A location of interest is identified, and corresponding spatial coordinate transformation is used. The integral measure is evaluated for each region giving a set of time series.
%The scale-space construction is repeated for each and the results are stacked into the SSST cube. 

\subsection{Numerical Treatment}
This section so far provides a rigorous approach to constructing a scale-space for particle datasets in reformed coordinates.
Following the definitions for the numerical implementation would, however, require the recomputation of the function $\Phi^\star$ using Eq.~\ref{eq:Phi_star}
for a range of scales. 
As summing over all the particles is the most demanding part of the computation, we can compute the values of the projected KDE only once, for a small value of $\sigma$ and obtain the higher scales by integral transformation (generalized convolution) with the projected kernels.
\begin{equation}\label{eq:integral_transform}
\Phi^\star_{\sigma + s}(t,u) = \int \gauss^\star_s(u, u_0)\cdot \Phi^\star_\sigma (t,u_0)\ \dd u_0
\end{equation}
Looking at the shapes of the projected kernels in Fig.~\ref{fig:both}, we conclude that they copy the Gaussian kernels, except for the boundary. 
With care, we can take the optimization even further and approximate the scale-space by convolving the function $\Phi^\star(t,u)$ with a simple Gaussian kernel:
\begin{equation}
    \Phi^\star_{\sigma+s}(t,u) \approx \gauss_s(u) \ast \Phi^\star_\sigma(t, u)
\end{equation}
This is an approximation only in the spatial case. 
For the temporal scale the Gaussian convolution is the exact result.

%It is also important to sample the axis $u$ densely enough with respect to the KDE width $\sigma$, such that the kernel spreads over enough sample points. \blue{(This should be obvious to anyone that has worked with discretizations.)}

%%%%%%%%%%%%%%%%%%%%%%%%%%%%%%%%%%%%%%%%%%
\section{Visualization \& Interaction Design}
%%%%%%%%%%%%%%%%%%%%%%%%%%%%%%%%%%%%%%%%%%
In this section, we present the use of scale-space splatting as a method to bring spatial information into aggregate measures in the analysis of molecular dynamics (MD) data.
We do not provide a replacement of well-proven tools, such as VIA-MD\cite{molva.20181102}, which are clearly useful for many purposes.
Instead, we demonstrate that considering the spatial context of the aggregate measures provides additional information, which can be exploited to build improved tools for visual analytics \blue{of MD simulations}.

The goal of our visual analytics \blue{solution} is to enable the exploration of multi-scale features \blue{using} the aggregate measures of trajectory-based spatio-temporal data in molecular dynamics.
The aim is not to promote the exploration of scale-space over the original simulation, but to enhance the analysis with a spatio-temporal scale-space approach.
The standard way of exploring such data is through 3D animated rendering of the spatial configuration, steered in time using an interactive time axis.
An integral measure, such as interaction energy, represented through a timeseries $f(t)$ using a line plot, is often used for navigating the time axis.
However, a simple line plot is not sufficient to capture the multiscale spatio-temporal behaviour.

We propose a multiple coordinated view solution for exploring multiscale features of the MD simulation. 
This is done by navigating and brushing in a scale-space representation linked to a 3D view showing the selected spatiotemporal intervals.
Users can select temporal ranges to loop over in an animated 3D view where the spatial extent is expressed in focus and context highlighting manner.
Efficient computation of the spatiotemporal scale-space of the integral measure using the scale-space splatting algorithm enables interactive exploration of scales even for large datasets.

The views include a 2D representation of the temporal scale and a 2D representation of the spatiotemporal slices of the scale-space space-time cube.
%scale view
In order to navigate the temporal scale, we propose a representation combining a line plot enhanced with a temporal scale-space image, which provides the context of different scales (see Fig.~\ref{fig:scale}).
%This way we combine the well-established concept of line plots with a new representation.
We provide interaction with the scale dimension and visual cues to support the exploration, as further detailed in Sec.~\ref{sec:scale-view}.
%We combine the line plot with the (temporal) scale-space image of the timeseries. 

In order to explore the spatial dependency of the integral measure at different scales, we render slices of the \ssst{} cube (Sec.~\ref{sec:ssst}). 
We do this by encoding the density field as a pixelmap, where the colour of each pixel encodes the values of the integral measure at corresponding spatial and temporal coordinates. This is described in Sec.~\ref{sec:slice-view}.
Both views share the temporal scale axis, which enables the exploration of the temporal scale in the scale view, where it is easier to grasp the effect of the reconstruction kernels.
As it is difficult to read off exact values from the density plots using colormaps, plotting of the slices of the slice view is also supported. 

The description of the linking and investigating the spatial configuration of the particles in a 3D rendering is described in the Sec.~\ref{sec:3d-view}.

An overview of our application can be seen in Fig.~\ref{fig:teaser}.
\blue{A demonstration of supported interactions can be found in the video in the supplementary materials.}
%\begin{figure}[th]
%\includegraphics[width=\columnwidth]{pictures/application.png}
%\caption{Overview of the solution.}
%\label{fig:overview}
%\end{figure} 
%

\subsection{The Scale-Space Space-Time Cube} \label{sec:ssst}
Bach et al.~\cite{bach2017space-time-cube} introduced a unified framework of space-time cubes. 
In their work they summarize the nomenclature for data manipulation and visualization of spatio-temporal data.
A conceptual image of a cube is introduced to unify the spatial and temporal axes.
An example would be stacking the frames of a movie into a box. The $x$ and $y$ coordinates correspond to the screen space coordinates, whilst the $z$ axis corresponds to time. 
Playing the movie then corresponds to slicing the box (cube) perpendicular to the $z$ axis.
Applying the space time cube to the movie data allows for exploring novel visualization techniques such as using non-orthogonal slicing, drilling, flattening (aggregating) and other data transformation techniques.

In order to manage our construction in a visualization, we introduce the concept of a \ssst{} (SSST) cube. An SSST cube builds on the concept of the space-time cube as described by Bach et al. but treats one of the axes as temporal scale and one as spatial scale.  
In our case, the cube has 4 dimensions: time, reformed space, temporal scale, spatial scale.
The temporal scale is constructed by convolving the data with Gaussian kernels along the temporal dimension. The spatial scale is constructed by convolving with Gaussian kernels along the new spatial axis.
We slice the cube in time-space planes in order to visualize spatiotemporal features.
We flatten the cube along the spatial axis, to obtain the standard time series representation of the integral measure as computed over the whole spatial domain, enhanced with the time-scale scale-space information.
We drill along the spatial axis for direct visualization of values at given time and scales.

\subsection{Scale View}\label{sec:scale-view}
\begin{figure}[btbp]
\includegraphics[width=\columnwidth]{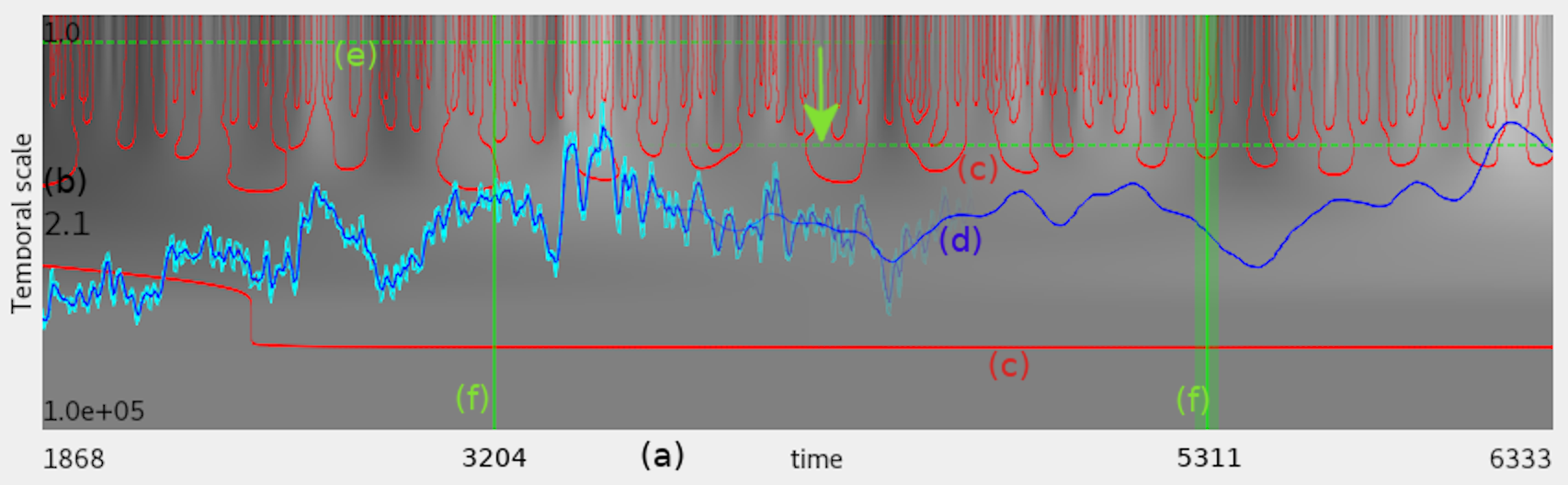}
\caption{\emph{The scale view} provides the investigation tool for the temporal scale of the data. 
The horizontal axis is time (a), the vertical axis is temporal scale (b). 
The background is the scale-space image with its second derivative level-set highlighted in red (c). 
The value of the timeseries at a selected scale is in blue (d) with the min-max interval in cyan. The change of the scale control (e) from value $\tau =2.1$ to $\tau = 16.1$ results in the disappearing of the min-max interval and a thicker window around the time pointer (f). The picture is a horizontal blending (left to right) of two screen-shots with the time indicator at different positions.}
\label{fig:scale}
\end{figure} 
We find that there are many visualization papers utilizing scale-space approaches, however, there is not much research done on direct visualization and interaction techniques for scale-space datasets. 
There are two examples, both using a colourmap to show a scale-space image. An example from statistics \cite{SiZer}, uses a scale-space image coloured based on statistical properties of the reconstructed timeseries to find prominent features.
%The visualization of the image itself is done by a grey-scale colourmap.
The second example is Pinus \cite{pinus}, which computes a hierarchical aggregation of the investigated timeseries. They employ a triangular-shape image representation of their scale-space, where pixels are coloured by the aggregated values.
%The screen size of their representation grows quadratically with the number of timesteps.

%For visualization of the timeseries reconstructed at a particular scale a line plot is commonly used issue with a long timeseries representation is the overplotting.
%Here we can look at the publications on multiscale time series visualizations using interactive multi-scale time axes \cite{Midgaard, TimeZoom, FacetZoom, Wakame, MultiscaleBehaviour}. 
%These mostly use three discrete scales such as Day--Month--Year. 
%There are also techniques using aggregation, such as BinX \cite{ BinX} and VizTree \cite{VizTree}. 
%For long timeseries data in work of Duran et al.\cite{Duran2018} the quartils together with the resampled version of the original timeseries are shown.

%
For showing the temporal scale-space of the timeseries $f(t)$, we adapt the direct colourmapping of the scale-space image, similar to both SiZer\cite{SiZer} and Pinus\cite{pinus}. 
Since we are dealing with long timeseries data and the representation in Pinus scales quadratically with the number of timesteps, we adopt the log-scaled scale axis from SiZer.
This way we end up with showing a long strip $n \times \log(n)$ pixels, rather than a huge $n \times n$ matrix, (the background image in Fig.~\ref{fig:scale}).

The log-scaling is justified by linearizing the effect of the blurring of the time-series.
A small change $\Delta\tau$ in the size	 of a large kernel $\tau >> \Delta\tau$ has very little effect on the resulting time series. 
On the other hand, the same change has much bigger effect for kernel size comparable to the change itself ($\Delta\tau \approx \tau$). 
Using the exponential function to explore \red{(sample) }the scale dimension keeps the relative change of the kernel size $\frac{\Delta\tau}{\tau}$ constant. 
Exponential sampling of the dimension leads to the log-scaled axis, ensuring that a fixed change anywhere along the re-scaled axis has a proportional effect on the appearance of the timeseries.
In practice, using a linearly scaled axis would result in the uninteresting, grey, bottom part of the image being much wider, up to the point of not having anything else in the picture. 

Since any colourmap itself is insufficient for reading off exact values of the timeseries,
%at given scale and time
we provide slicing of the image, similar to the approach in Pinus \cite{pinus}.
The only difference is that we plot the timeseries on top of the rectangular scale-space image (Fig.~\ref{fig:scale}d).
The selected scale is indicated by a horizontal dashed line (Fig.~\ref{fig:scale}e) and the exact value of scale is displayed on the left (Fig.~\ref{fig:scale}b).
Due to the finite resolution of the display, it is impossible to show the reconstructed function for some combinations of zoom and scale.
This effect is well known in the field and can be addressed for example by using a band graph \cite{Aigner:2011:VTD} as in BinX \cite{BinX} and Duran \cite{Duran2018}.

Sliding the green dashed line up and down provides means for exploring the temporal scale space.
We found that the scale-space image in the background of the line plot successfully highlights prominent regions in the data, but it provides only a vague indication of interesting scales. 
We therefore include the red curves (Fig.~\ref{fig:scale}c), which are the contours of zero crossings of the second derivative computed according to Eq.~\ref{eq:contours}. 
They provide a visual cue for regions of peaks and valleys in the data, together with corresponding scale significance. 
The ``deeper'' the contour reaches, the more prominent the given peak or valley is. The feature disappears from the line plot completely when the selected scale does not cross the contour. 

In order to interact with the time axis, the selected time indicator is drawn as a vertical green line (Fig.~\ref{fig:scale}e), which can be dragged along the time axis and changes the currently selected time step. 
Since the selected time step at the current time scale is the result of averaging over a neighbourhood
%(for both the time series and the \ssst{}) 
the indicator is enhanced with a semi-transparent green box showing the size of this neighbourhood.
It is drawn around the vertical green line (Fig.~\ref{fig:scale}f) showing the $(t - \tau, t + \tau)$ interval corresponding to the width of the Gaussian kernel of the currently selected scale. 
\subsection{Slice View}\label{sec:slice-view}
\begin{figure}[b!]
\includegraphics[width=\columnwidth]{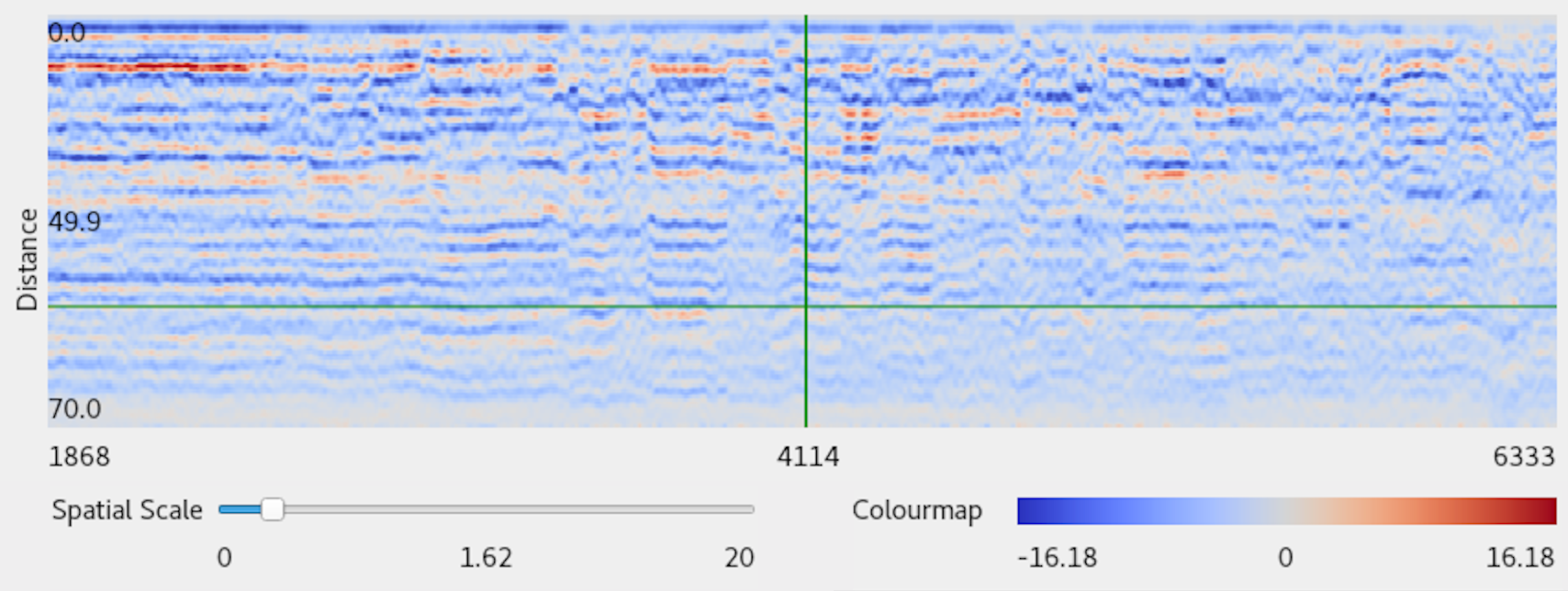}
\caption{The slice view: a pixelmap showing a spatio-temporal slice of the SSST cube with horizontal time axis and vertical spatial axis. The green cross consists of a horizontal line for navigation in the spatial axis and a vertical line for navigation along the time axis. A divergent colourmap is used to encode the values. The spatial scale is set to $\sigma = 1.62$ \AA{} and temporal scale is shared with the scale view $\tau = 2.1$. }
\label{fig:slice_view}
\end{figure} 
The slice view provides the spatiotemporal overview of the simulation with respect to the chosen integral measure.
It does its job by conveying the information contained in the \ssst{} cube.
The cube is a four dimensional density field which is traditionally visualized with 1D \cite{Sliceplorer} and 2D \cite{Hyperslice} slices.
The idea is to incorporate the spatial information into the lineplot representation of the timeseries. 
To preserve coherence between the two representations, we keep the temporal axis of the line plot and use the second axis for the reformed spatial dimension.
Since we can no longer use the second axis for encoding the values, as in the line plot, we utilize a pixelmap equipped with an appropriate colourmap.
The colour of each pixel is determined by the value in the \ssst{} at given spatial and temporal scales, with the x coordinate corresponding to the time axis and the y coordinate corresponding to the reformed spatial axis.
To support the exploration of exact values in the pixelmap we enable plotting of the spatial dependency at the given timestep and the particular scales in a separate view (Fig.~\ref{fig:teaser} d).
%For the purpose of visualizing the positive-negative values, specific to our dataset and the chosen measure, we used Moreland's divergent colourmap~\cite{Moreland2009}. 

%This spatiotemporal slice of the scale-space cube can only be represented at particular levels of spatial and temporal scales.
When the frequencies in the data are higher than what is possible to capture on the screen, the pixelmap suffers from the same overdraw as the line plot.
Since there is not an easy solution for including the overplotted ranges in the pixelmap, we rely on the joint properties of the scale view and the slice view. 
The slice shares its values with the timeseries representation -- the values in the slice aggregated over the spatial domain yield the timeseries -- they will also roughly share the same frequencies in the data. 
We exploit this correspondence by sharing the temporal scale axis between the two views.
It is thus possible to perform the exploration of the temporal scale in the scale view and propagate the interaction to the slice view.
The scale view thus serves as a proxy for exploration of the temporal scale of the slice view with explicit indication of the effect of the overdraw.

A similar scale exploration tool should, in principle, be available also for the spatial scale, where the aggregate of the slice along the temporal axis would guide the exploration.   
In our case, where the temporal axis is much larger than the spatial axis, we found it sufficient to enable the spatial scale exploration by a slider and investigating the effect of the scale selection in a simple line plots of values of the cube along the spatial dimension for a selected time (see Fig~\ref{fig:teaser} d).
An advanced tool, such as the scale view, is necessary for exploration of a scale axis when interacting with various level of zoom.

The second link between the slice view and the scale view is realized through the shared temporal scale axis, which supports panning and zooming. 
%The interaction with the dashed line in the scale view is used for navigating the temporal scale in the slice view.
%The direct interaction in the slice view supports selection of the particular spatio-temporal point indicated by the green cross in Fig.~\ref{fig:slice_view}.
The green cross in Fig.~\ref{fig:slice_view} indicates the temporal and spatial selection used for the 3D view. 
%
%One needs to remember that we are trying to visualize a density values along a reformed coordinate, which may lead to misinterpretations.
%The projection of the scale-space onto the reformed coordinate is going to scale the values by the Jakobian of the transformation.
%
%
\subsection{3D view}\label{sec:3d-view}
\begin{figure}[th]
\includegraphics[width=\columnwidth]{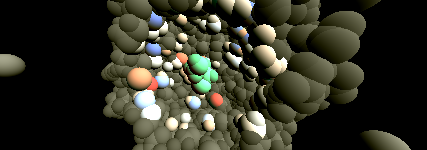}
\caption{An example of the 3D view from the molecular dynamics simulation showing the peeling of spatial scales around the point of interest. The particles in the selected region are coloured based on their contribution to the integral measure. The atoms are coloured based on their contribution to the integral measure and the colourmap is shared with the slice view. The ligand representing the point of interest is in green.}
\label{fig:3d_view}
\end{figure}

There is a body of work on 3D visualizations of particle data in MD, SPH, and also other fields \cite{alharbi2017molecular,SPH-Rendering}.
\blue{Since 3D representations are not central in our research, we default to the simplest, useful, particle representation by shaded spheres at particle positions with user-adjustable sizes.}
%All the values in the SSST cube can be resolved back to particular space and time localization.
We render the configuration of the particles at the time step selected in the shared temporal axis in slice and scale view. 
The way of focusing on a given spatial scale is by hiding inner or outer regions with respect to the decomposition. 
The spheres in the focused region are coloured with the same colourmap as in the slice view, with the values based on the contribution of each particle to the integral measure.
By this, the user is able to pinpoint a particular interaction both spatially and temporally. An example of peeling the concentric spheres from around the point of interest in a protein--ligand simulation can be found in Fig.~\ref{fig:3d_view}.

\subsection{Implementation}
The implementation consists of two parts, a pre-processing module and an interactive visual analytics tool.
Both the data pre-processing and the visual analytics tool are implemented in Python.
For a given measure and a reformation, the pre-processing consists of filling the data into one spatiotemporal slice of the \ssst{} cube at the highest resolution.
For this the whole simulation is processed frame by frame, where each particle is transformed into the new coordinate system and its contribution is projected onto the chosen axis.
The timeseries is obtained by aggregating the slice along the spatial axis.
The scale-space image of the timeseries is constructed by repeated convolutions with Gaussian kernels and the zero crossings are traced with sub-pixel precision.

All the other slices of the \ssst{} cube are computed on demand by convolution with Gaussian kernels. 
As demonstrated in Sec.~\ref{sec:reform}, the projected kernels are also Gaussians, except for when they overlap with the boundary $r \rightarrow 0$. 
Since in our data all the values near the $r = 0$ are zero (see Fig.~\ref{fig:teaser}, top right), we do not explicitly implement the boundary condition as described in Eq.~\ref{eq:integral_transform}.

The user interface has been designed using bindings for the Qt framework.
It runs interactively with the only demanding parts being convolving the matrix in the slice view with 1D Gaussian filters, recomputed on the scale change. 
This could be easily optimized by convolving the matrix on the GPU.
%
%
%%%%%%%%%%%%%%%%%%%%%%%%%%%%%%%%%%%%%%%%%%
%%%%%%%%%%%%%%%%%%%%%%%%%%%%%%%%%%%%%%%%%%
\section{Demonstration}
%%%%%%%%%%%%%%%%%%%%%%%%%%%%%%%%%%%%%%%%%%
%%%%%%%%%%%%%%%%%%%%%%%%%%%%%%%%%%%%%%%%%%

In the following, we describe the use of the new method in our proof-of-concept application. 
We worked on two datasets, each in collaboration with a domain expert: one protein engineer and one expert in computational biology with background in quantum chemistry.
The data consists of the trajectories of the atoms of a protein and of a smaller molecule called a ligand. 
One of the integral measures commonly used for the analysis of MD simulations is electrostatic interaction energy.
 
The electrostatic interaction energy is computed from the pairwise interaction of the atoms of the ligand and the protein by the formula:
\[
F_{el} 
= \sum_i \underbrace{
	\sum_k \frac{1}{4\pi\varepsilon_0} \frac{q_i q_k}{||\vec{p}_k - \vec{p}_i||}
}_{\mathcal{F}_i(\vec{p}_i)} 
\]
where $i$ is indexing the atoms of the protein, $k$ is indexing the atoms of the ligand, $\vec{p}$ is the atom position, and $q$ is the atom partial charge. The letter $F$ does not stand for force, but for the integral measure, in accordance with the definition in Eq.~\ref{eq:state_fnc}.

Since the position of the ligand is of importance in the analysis, we identify the position of the ligand as the reference point and present an analysis of the data using the spherical transformation. \blue{The details of the transformation are  described in Sec.~3 of the supplementary material.
Further results obtained using cylindrical and orthogonal transformations are demonstrated in Sec.~4-5 in the supplementary material.}

\subsection{Three Ligands}

The first dataset we studied contains 50k timesteps of an MD simulation of a protein with 4650 atoms and three ligands, each consisting of 12 atoms. 
\red{The spatial domain of the simulation is a cube with a side length of 100\,\AA. The atom radii range from 1.2\,\AA{} to 3.48\,\AA{} with an average~of~1.97\,\AA{}.}
The research question addressed with the simulation is the analysis of the ligands' propagation to the active site. \blue{The data come from the research published in the paper by Marques et al. \cite{Marques2017}.}

Exploring the spatio-temporal scale-space of the integral measure improves navigation through the simulation.
Comparing the timeseries representation of the electrostatic interaction energy with the slice view, the slice view provides an immediate indication of the intervals when the ligand is interacting with the protein and when not (see Fig.~\ref{fig:teaser} a).
The visualization also shows, that there is no interaction past the 25 \AA{} distance, which is important information for correctly setting up the simulation and is an object of study on its own \cite{fadrna2005long}.
%This is confirmed by the domain expert and explained by the ligands having zero nett charge.
Zooming on the 0 -- 25 \AA{} interval is performed at the data level, but interactive spatial navigation would be necessary for a final-product application.
%Looking at the spatially zoomed-in overview, we notice that the gaps in the pixelmap correspond to ligand floating freely around and the non-zero data correspond to bound states.
%The timeseries representation of the integral measure, once again, does not provide sufficient guidance for exploration of the data.
When investigating ligand no. 3 at appropriate spatial and temporal scales, there are three distinct features that draw instant attention.
The first one are the empty intervals towards the beginning of the simulation, where the ligand is too far away from the protein (Fig.~\ref{fig:teaser} a).
The second one starts with the ligand trying to escape the tunnel and as it approaches the surface it is pushed back into the tunnel. A second ligand appears at the entrance, locking the ligand no. 3 in the active site (Fig.~\ref{fig:teaser} b). 
The third interesting region occurs at times when all three ligands are present near the same tunnel opening (see Fig.~\ref{fig:teaser} c).
Comparing these features to the very weak indications in the timeseries representation, our proposed visualization makes the identification of relevant features in the data more apparent.
\subsection{Five Ligands}
%%%%%%%%%%%%%%%%%%%%%%%%%%%%%%%%%%%%%%%
We have obtained an MD simulation dataset from our collaborators consisting of 5 ligands (26 atoms each) and a protein of circa 16 thousand atoms, with the duration of 10\,000 timesteps. 
The research question that led to the computation of this simulation was to study the ligand's dynamics in the binding pockets over time.
%The domain expert had already spent hours understanding the dataset.
We received the dataset with a conclusion: \enquote{There is nothing [to be seen] in the electrostatic energy. It has to be coupled with other measures.}.

\begin{figure}[tb!]
%\begin{subfigure}[t]{.5\textwidth}
\includegraphics[width=\columnwidth]{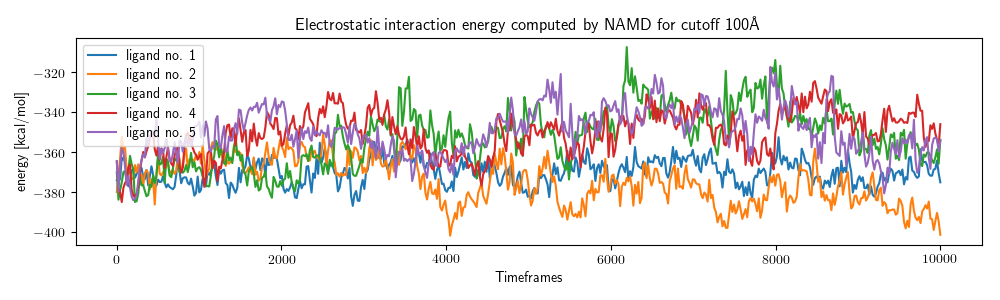}
\caption{Comparison of the electrostatic interaction energies for the ligands no. 1 - 5. There is no clear deviation of the energy values for the two cases that undergo the change.}
\label{fig:energyAG}
%\end{subfigure}
\end{figure}

The plot of the energies can be found in Figure~\ref{fig:energyAG}.
Indeed, the overview plot does not separate the ligands into distinct categories, nor does it segment the time axis into different modes.
%The following paragraph describes the investigation done by the authors of this paper using our tool on the provided dataset. 

Loading the pre-computed data into the tool we find that 10 000 timesteps are causing overdraw on the time series representation and there is little to be seen in the initial scale slice.
Interacting with the temporal scale reveals that in fullscreen on our 1920px wide monitor the $\tau$ of 9 timesteps results in a smooth plotting of the energy function.
Due to the joint treatment of the temporal scale axis we also automatically obtain a good setting for the slice view.
Hence, a preliminary vertical pattern appears also in the slice view.
Interacting with the spatial scale, after some attempts we identify that above 4 \AA{} a vertical pattern is dominant in the data and the horizontal stripes get blurred out.
The scale size of 2 \AA{} preserves the structure and enhances the contrast.
%We can compare the effect of smoothing along the spatial domain in the line plots of the data.
%We can improve the contrast further by scaling the colourmap. 
%The factor of 1.5 brings out the most details.
\begin{figure}[bt!]
    \centering
    \includegraphics[width=\columnwidth]{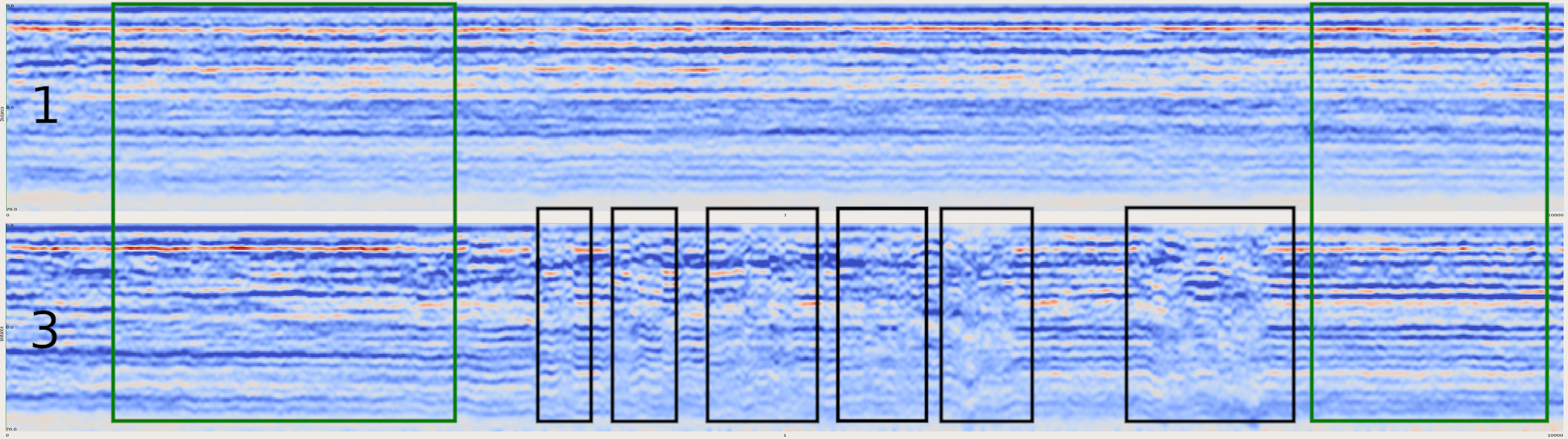}
    \caption{The comparison of the slice views for ligands no.1 and no.3 providing an overview of the simulation. The ligand no.1 exhibits stable behaviour during the whole simulation as opposed to the ligand no.3 which undergoes several changes, framed in black. Both ligands show similar behaviours towards the beginning and the end of the simulation.}
    \label{fig:comparison}
\end{figure}
Comparing the overview visualizations for all the ligands we find that ligands no. 1, 2 and 4 exhibit stable behaviours throughout the simulation, whereas ligands no. 3 and 5 experience various changes.
%Having tuned our visualization, we compare the overviews of the whole simulation for different ligands.
%Switching over to ligand no. 2, we find the same spatio-temporal pattern in the slice view.
%Switching to ligand no. 3, we observe the same pattern in the beginning and towards the end of the simulation, however, there are several intervals where the pattern breaks (see Fig.~\ref{fig:comparison}). 
%The ligand no. 4 shows a stable pattern with slight variations over time.
%The ligand no. 5 shows the same pattern with fewer but stronger changes.
%Zooming in on a change with interval length of circa 500 timesteps shows low frequency signal both in scale view and spatial view. 
%Choosing the temporal scale corresponding to the original data brings out very noisy pattern.
%Adjusting the temporal scale according to the contours such that many contours are crossed by the scale indicator, but not all of them (smallest ones) brings out the structure due to temporal coherence once again.

\begin{figure*}[tbhp]
    \centering
    \includegraphics[width=\textwidth]{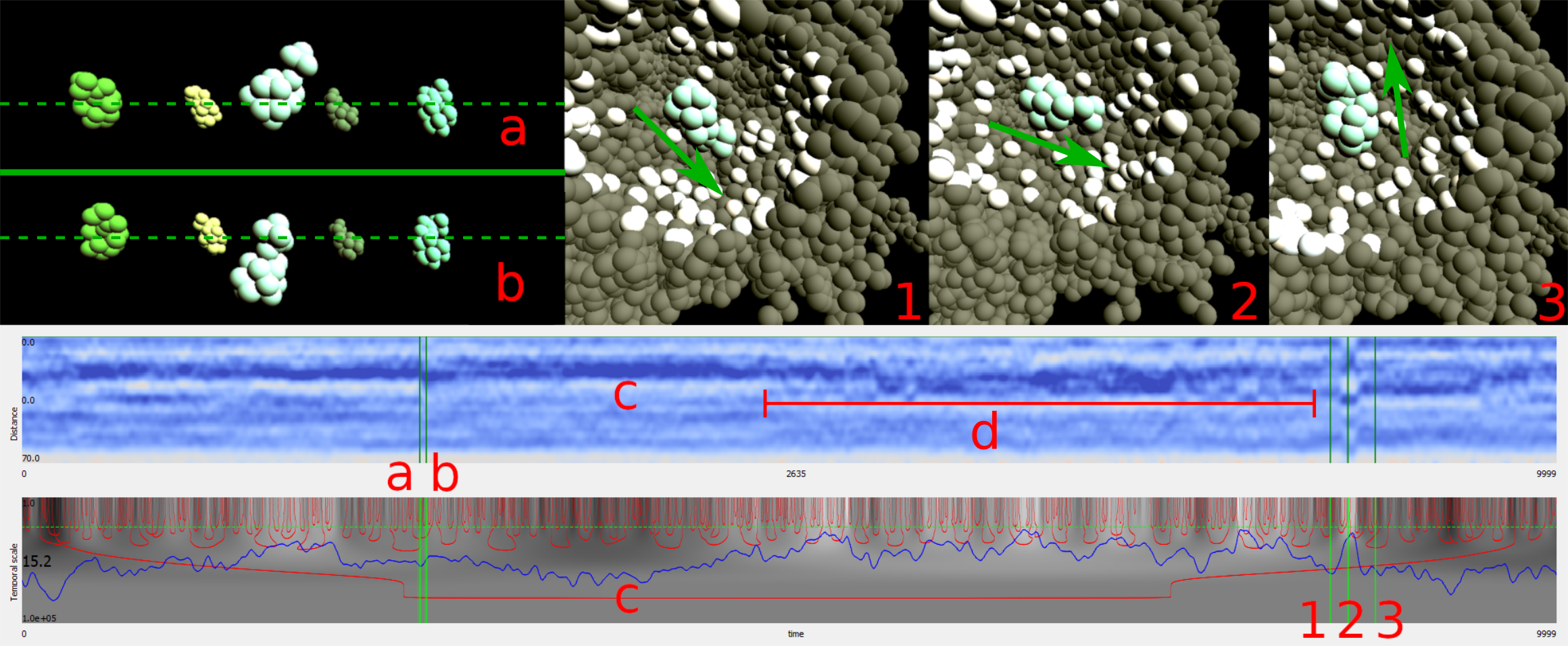}
    \caption{The analysis of behaviour of ligand no. 5 at large spatial scale. The two most prominent features in the slice view (middle) are the shear towards the beginning of the simulation (a,b) and the abrupt modality change towards the end (1,2,3). The shear is caused by movement of the ligand in the binding pocket. Protein is not shown to avoid occlusion, instead the movement is referenced by the  plane that ligands lie in. The shear follows by energy minima which is largely contributed by the protein atoms located at a distance between 16.6 - 22.2 \AA{} (c). The change of modality (1,2,3) results after a period of unstable behaviour (d) and is resolved as ligand's rotation (the arrows indicate the ligand's orientation).}
    \label{fig:lig-5}
\end{figure*}

The cross-scale exploration enables the identification of prominent features in the data.
For example, applying a considerable amount of blurring in the spatial domain ($\sigma = 5.86$ \AA{}), and a relatively low amount in the temporal domain ($\tau = 10.7$), we see spatially significant changes that possibly happen over short time periods.
The effect of these settings on the slice view for the ligand no. 5 can be seen in Fig.~\ref{fig:lig-5} (the mostly blue pixelmap).
Two significant events are apparent in the scale view. 
The first change in the pattern is a shear occurring at timestep 2600 (Fig.~\ref{fig:lig-5}a,b). 
Investigating the 3D configuration shows a big movement of the ligand inside of the binding pocket.
This event is not apparent in the energy timeseries, however, it allowed the ligand to reach a better position as the value of the energy reaches its minimum after this change (\ref{fig:lig-5} c).
The biggest contribution to this energy minimum comes from atoms located at distances between 16.6 and 22.2 \AA{} from the ligand's centre.

The second significant event happens towards the end of the simulation, where the original pattern is interrupted and it is paired with a spike in the energy timeseries (Fig.~\ref{fig:lig-5} 1,2,3). 
The snapshots of the ligand before, during and after the change show the ligand changing its orientation.
%With the help of the domain expert we realise that the different modes in the scale view correspond to different orientations of the ligand in the pocket observed in the 3D view.

The following can be observed from the example described above.
The descriptive power of the fully aggregated integral measure was not sufficient to distinguish the interesting behaviour. 
Computing the spatial dependency in full resolution and simply showing the data does not suffice for finding patterns in the data.
Exploring the scale-space representation enabled rapid identification of prominent features, which lead to spatial and temporal localization of interesting behaviour.
%Having a navigation tool for the temporal scale axis helps navigating the scales as opposed to a blind sampling of the spatial scale. 
%ith our tool we were able to identify the temporal and spatial localization of the events -- when and where the changes have taken place. 
This example demonstrates the analytical power of accessing spatial information of an integral measure.
The domain expert confirmed that the timeseries representation does not provide sufficient guidance in this particular case.
On the other hand, being able to spatially resolve the respective contributions to the timeseries at correct scales, we are able to gain more insight from exactly the same data.
%e can also see, that a correct treatment of scales is fundamental for meaningful encodings of the values and a sufficient analysis.
We have thus demonstrated that there is more information accessible through the integral measure when the spatial dependency is also included and relevant scales are considered. 
%\begin{figure}
%\includegraphics[width=\textwidth]{•}
%\end{figure}
\blue{\subsection{Domain Expert Feedback}}
\blue{The presented use case scenarios were worked out in cooperation with two domain experts in order to validate our results. 
To evaluate the usability of the prototype tool, we have conducted a 45 minutes demonstration followed by an interview with one of the domain experts. 
The domain expert answered questions to ensure his understanding of the application and was tasked to guide a short exploration of an MD simulation, while the authors were interacting with the application. 
}

\blue{
For the \enquote{three ligands} dataset two spherical projections were available, one centred at the ligand and one centred at the protein. 
The expert chose to work in the spherical projection centred at the protein and pointed out that the slice view provides a good overview of the simulation.
He navigated to a time interval corresponding to an extremum in the energy and observed that for a detailed analysis more context in the 3D view, such as amino-acid/atoms labels and different 3D representations would be helpful.
The strong point of the slice view was that the spatial orientation of the ligand with respect to the protein could be deduced.
He also mentioned that \enquote{It would be interesting to compare the Van der Waals and electrostatic using this view.}
While the Van der Waals energy is not at the moment supported, it is an example of an integral measure that can be easily incorporated in the future.
He found the treatment of the scales appropriate as there are many different time scales in the simulation and one could easily examine the persistence of features across scales.
With respect to the usability he mentioned, that \enquote{It would, of course, require some training, but it does not seem more difficult than other analytical software available.}.
}

\subsection{Performance}
The pre-computation of the spatial decomposition along the ligand trajectory is the most demanding part.
For the decomposition of 5 ligand trajectories against a protein with 16 thousand atoms and 10\,000 time-frames the pre-computation took 20 minutes per ligand in a single core application on a system with following parameters:~ CPU: Intel(R) Core(TM) i7-7700K @ 4.20GHz;~ RAM: 2 x 16GiB DIMM DDR4 2400 MHz;~ GPU: GeForce GTX 1080, 8 GB GDDR5X;~ Storage: 500GB SSD Samsung 860 Evo.

A full ligand--protein electrostatic interaction for a ligand consisting of 26 atoms took 6.5 hours.
We acknowledge that the implementation is not optimized and better tools for computing these values are available~\cite{NAMD}. The domain experts do not find the computational time to be an issue, since our precomputation time is negligible with comparison to the computational time of the simulation.
The result of the precomputation is scalar 2D array of $n \times m$ values, which is easily dealt with both on CPU and GPU.

\section{Discussion and Limitations}
We have demonstrated that the simultaneous treatment of spatial and temporal scales in the analysis of MD simulation can lead to an improved analysis.
Of course, this information does not have to be accessed in every analysis, namely when the timeseries representation sufficiently captures the studied phenomena.
Being able to access the additional information is vital when other approaches fail. 

\blue{
The reason that our method works better in these cases lies in enabling exploration of the spatial structure by efficient computation of scale-space. 
Where previous approaches would be able to reconstruct the spatial representation of the integral measure, they would do so at a fixed resolution \cite{NAMD}, without an interactive visual support.
We, on the other hand, enable a multiscale exploration of both spatial and temporal dependencies of the integral measures, which allows more flexibility in the analysis.}
\red{
One can object that in the end the method boils down to Gaussian blurring of traditional visualization techniques.
This would be a valid observation with respect to the resulting implementation. 
Applying Gaussian blur to data is, indeed, the core of scale-space techniques.
However, the technique is a result of careful and well-founded theory and not just ad hoc application of image filtering. 
\begin{flushright}
[This paragraph was removed, because it is not constructive. Instead, the requested comparison was included (blue above).]
\end{flushright}
}

The biggest obstacle for a widespread use of the method is a steep learning curve regarding the scale-space.
In particular, we had hard time explaining the scale-space basics and the mechanics of the scale view to the domain experts.
Yet, they found our demonstration of the analysis powerful and the results relevant.
The steep learning curve of scale-spaces needs to be balanced with an attractive, well-designed, user-friendly interface that can bridge the gap with ease of use over years of knowledge in calculus and differential equations.
\blue{This would ideally be combined in one of existing MD analysis tools, since the domain experts rely on more than one approach in a successful analysis.}

%%%%%%%%%%%%%%%%%%%%%%%%%%%%%%%%%%%%%%%%%%
%%%%%%%%%%%%%%%%%%%%%%%%%%%%%%%%%%%%%%%%%%
\section{Conclusion and Future Work}
%%%%%%%%%%%%%%%%%%%%%%%%%%%%%%%%%%%%%%%%%%
%%%%%%%%%%%%%%%%%%%%%%%%%%%%%%%%%%%%%%%%%%

In this work, we presented a method for investigating the spatial dependence of integral measures of particle data across temporal and spatial scales. 
Based on the theoretical framework we have implemented a sample demonstration application showcasing the potential benefits of simultaneous spatial and temporal scale exploration of trajectory-based particle data in molecular dynamics. 
We have shown that this kind of analysis is suitable for the exploration of one-dimensional spatial dependencies in aggregate measures. 
Consequently, such exploration can be a valuable tool for real world applications, helping in understanding long time simulations.
\blue{As the core of the method depends only on trajectory based simulation, equipped with an integral measure, exhibiting a simple symmetry, we expect that it could be applied to other fields outside of molecular dynamics. This will be subjected to further research.}

There are several research directions open for future work.
%One might be consider hiding the scale-space from the user and only using the framework for optimal reconstruction driven by the screen resolution.
A (semi-) automatic scale identification could be investigated for pre-selection of relevant scales to enhance the user experience.
Furthermore, the unification of spatial and temporal scales might allow for automatic feature detection in the spatiotemporal scale-space similar to the works of Laptev \cite{Laptev2005} and Lindeberg \cite{Lindeberg1993, Lindeberg2011}, however, this might be strictly domain or even data dependent.
There is also a possibility for adapting the framework to non-uniform scale-space constructions, similar to the edge enhancing diffusion of Perona and Malik~\cite{PeronaMalik}.

Also more visualization techniques could be adapted for the \ssst{} cube, namely  non-planar slicing -- adaptively changing the reconstruction precision based on an importance function.
Furthermore, promoting the scale-space as the prime object of interest and using direct volume rendering, or iso-surface extraction to trace its structure can be of interest. 
This might combine greatly with the Gaussian derivatives of the data.

%Last but not least, the domain experts would 

%Lastly the next step could be to focus on more than one dimensional spatial dependency and generalizing to vector fields. 

\acknowledgments{
We would like to thank our collaborators: Dr. S\'{e}rgio Marques from Loschmidt Laboratories and Hanif Muhammad Khan from Computational Biology Unit at University of Bergen, for providing us with the data and their invaluable feedback.
}

\clearpage

\bibliographystyle{abbrv-doi-hyperref}
\bibliography{manuscript}

\end{document}